# Quantum Computing Simulation Optimizations and Operational Errors on Various 2-qubit Multiplier Circuits

Eric Hsu
Department of Computer Science
Cornell University
eth5@cornell.edu
August 15, 2002

**Abstract**
Since simulating quantum computers requires exponentially more classical resources, efficient algorithms are extremely helpful. We analyze algorithms that create single qubit and specific controlled qubit matrix representations of gates.
Additionally, we use the simulator to investigate errors based on different probability distributions and to investigate the robustness of different 2-qubit multiplier circuits in the presence of operational errors.

## 1. Introduction

The theory behind quantum computing and information has advanced well beyond any physical implementation. Many theoretical performance bounds and capabilities of complex algorithms have been established. However, scientists have developed simple physical realizations of quantum computers at best. Quantum computer simulation will help bridge the disparity between high-level algorithms and quantum computer construction. Eventually we hope to optimize the layout of qubits and other physical components analogous to bits, circuits, and gates in classical computers.
Since simulating quantum information requires exponentially more processing power and resources on classical computers, we need efficient algorithms. We developed algorithms, which create single qubit gates and controlled qubit gates, that are moderately faster than the algorithms that other quantum simulators use. The algorithms were developed and analyzed in Matlab.
In addition, we used our simulator to explore elementary quantum circuits and their resistance to operational errors. Specifically, we are researching non-Gaussian distributed errors on density matrices and the robustness of various 2-qubit multiplier circuits.

## 2. Algorithm Comparison

### 2.1 Timing and Simulation Environment

The algorithms were written in Matlab version 6.1.0.450 release 12.1 for Linux. We used Red Hat Linux release 7.1 Kernel 2.4.9-31 on an i686. The test system has a 1.7 Ghz Pentium 4 and 1 Gb of RAM. We used Matlab's tic/toc commands to time the algorithms, and we used Matlab's Statistic Toolbox 4 to generate random numbers and

2conduct the ANOVA statistical tests. Also, since quantum algorithms generally take advantage of large amounts of entanglement, we stored our density matrices as full matrices.

## 2.2 Single Qubit Gates

To save space at the expense of operation time, we can generate a single qubit gate, S, from a 2×2 matrix U which represents the operation that we want to perform on a specific qubit. Various quantum computer simulators create matrix representations of single qubit gates over the entire n qubit register by using the following method: $S = (\otimes^{i-1}_{k=1} I) \otimes U \otimes (\otimes^{n}_{k=i+1} I)$ where n is the number of qubits, U is the operation that we want to apply to the $i^{th}$ qubit, and I is a 2×2 identity matrix. We call the $i^{th}$ qubit the target qubit (target = i). This equation translates into the following Matlab code. In the code, `kron(x,y)` computes the Kronecker tensor product of x and y, and `speye(m,n)` creates a m×n sparse identity matrix.

```
if (target == n)
        gate = kron( speye(2^(n-1), 2^(n-1) ), U  );

elseif (target == 1)
        gate = kron( U, speye(2^(n-1), 2^(n-1)));

else
        gate = kron( speye(2^(target - 1), 2^(target - 1) ), U  );
        gate = kron( gate, speye(2^(n - target), 2^(n - target)));

end
```

Figure 1

This algorithm is equivalent to creating a $2^{i-1} \times 2^{i-1}$ identity matrix, tensoring that identity matrix with the 2×2 matrix U, and then tensoring the result with a $2^{n-i} \times 2^{n-i}$ identity matrix. Note that we store gates as sparse matrices which only store and operate on nonzero values. In the worst case, that is when i is the $(n/2)^{th}$ qubit, the algorithm executes $2^{(n+1)}$ multiplications and has 2 function calls. When the target qubit, i, is either the $1^{st}$ or the $n^{th}$ qubit, we also need $2^{(n+1)}$ multiplications but we only execute 1 function call.

In practice, when the target qubit was the $(n/2)^{th}$ qubit, the algorithm took much longer compared to when the target was the $n^{th}$ qubit. But we found that targeting the $1^{st}$ qubit was actually slower than targeting the $(n/2)^{th}$ qubit. There are many reasons why this slowdown occurs. When we target the $1^{st}$ qubit, we take the Kronecker tensor product of a small matrix and a large matrix. As a result, the Matlab code that implements `kron(x,y)` spends more time doing row indexing operations. Conversely, taking the Kronecker tensor product of a large matrix by a small matrix involves more column indexing operations. Matlab stores matrices in column form, so row indexing operations are slower than column indexing operations.

Table 1: Timing for algorithm in Figure 1

| Number of qubits | Target = 1st qubit (sec) | Target = middle qubit n/2 (sec) | Target = last qubit n (sec) |
|---|---:|---:|---:|
| 14 | 2.97 | 2.76 | 1.77 |
| 15 | 6.01 | 5.72 | 3.93 |
| 16 | 12.3 | 11 | 7.69 |
| 17 | 25.3 | 24 | 14.9 |
| 18 | 62.5 | 50 | 30.0 |
| 19 | 138 | 107 | 75.7 |
| 20 | 285 | 213 | 151 |
| 21 | 505 | 464 | 310 |
| 22 | 1110 | 1040 | 637 |
| 23 | 2230 | 2100 | 1280 |

Note: Times are a sum of running the algorithm 50 times

We can also create a single qubit gate using the algorithm in Figure 2:

```
dims = n - target;
length = 2^(dims) ;
upleft   = U(1,1)* speye(length, length);
upright  = U(1,2)* speye(length, length);
downleft = U(2,1)* speye(length, length);
downright= U(2,2)* speye(length, length);
gate = [upleft, upright; downleft, downright];
for i=2:target
      dims = dims+1;
      length = 2^(dims);
      none = sparse(length, length);
      gate = [gate, none; none, gate];
end
```

Figure 2

If we use the number of multiplications as our metric for computational complexity, the worst case occurs when the target qubit is the 1st qubit. In the worst case, this algorithm requires $O(2^{(n+1)})$ multiplications. The algorithm requires only $O(4)$, multiplications in the best case when the target is the $n^{th}$ qubit. Time consuming Kronecker tensor products and multiplications are replaced by matrix concatenation, and we do $\theta(i)$ matrix concatenations where i is the target qubit.

In practice, this algorithm runs fastest when the target qubit is around the $(n/2)^{th}$ qubit. In this case, we balance the work between fewer multiplications, $O(2^{(n/2)})$, and less concatenation, $\theta(i/2)$. When the target is close to the $n^{th}$ qubit, the algorithm creates the matrix solely by matrix concatenation, and performance suffers slightly. When the target is close to the 1st qubit, we only cut performance time by a factor of 4 in comparison to the algorithm in Figure 1. The speed improvement arises because we divide the problem into 4 equal smaller problems. We present a comparison of the algorithms in Figure 1 and Figure 2 in Table 6 on page 13. It is clear that the algorithm in Figure 2 for creating single qubit gates is significantly faster than the method described in Figure 1.

Table 2: Times for algorithm in Figure 2

| Number of Qubits | Target Qubits | | | | |
|---|---|---|---|---|---|
| | $1^{st}$ | $5^{th}$ | $(n/2)^{th}$ | $(n-5)^{th}$ | $n^{th}$ |
| 14 | 0.129 | 0.955 | 0.963 | 1.19 | 0.106 |
| 15 | 2.88 | 1.92 | 1.71 | 1.74 | 2.11 |
| 16 | 4.77 | 3.14 | 2.87 | 2.79 | 3.74 |
| 17 | 9.31 | 6.56 | 5.87 | 5.57 | 7.60 |
| 18 | 17.5 | 12.8 | 11.5 | 11.5 | 14.1 |
| 19 | 35.8 | 25.1 | 22.8 | 24.6 | 29.1 |
| 20 | 73.3 | 48.9 | 44.7 | 50.4 | 55.4 |
| 21 | 147 | 95.5 | 83.4 | 97.3 | 105 |
| 22 | 306 | 190 | 170 | 198 | 212 |
| 23 | 620 | 445 | 360 | 372 | 421 |

Note: Times are a sum of running the algorithm 50 times

## 2.3 Controlled qubit gates operations

Additionally, we have created various controlled qubit gate algorithms that simulate controlled gate operations on the density operators without multiplying the density operators by matrices that represent the gate operation. Matrix representations of these gates are permutation matrices. We have developed algorithms for the controlled not, Toffoli, and Fredkin gates.

### 2.3.1 Density Matrix

To better explain our algorithm, we describe the density matrix. Note that we work with standard computational basis states. The matrices have |00...00> <00...00| in the upper left hand corner and |11...11><11...11| in the lower right hand corner. Below we explain in detail a 3 qubit system represented by a 8×8 matrix.

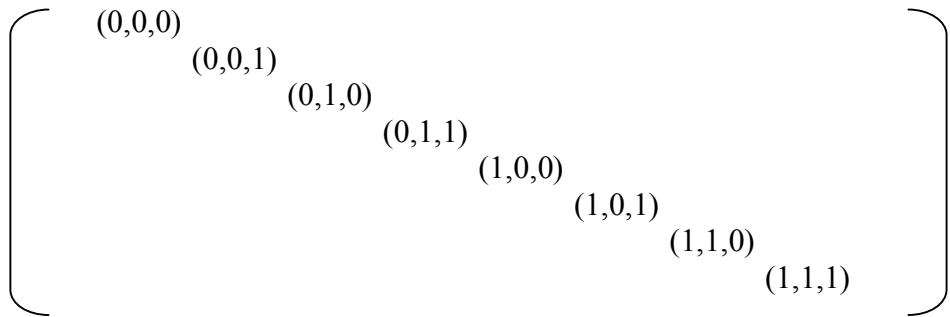

Figure 3: Layout of a Density Matrix representing a 3 Qubit System

We say that the left most bit in each tuple represents the first qubit, the second bit represents the second qubit,… and the right most bit in the tuple represents the last qubit, (i.e. $1^{st}, 2^{nd}, 3^{rd},…,n^{th}$). For convenience, we also number the tuples. (0,0,0) corresponds to the $1^{st}$ row/column, (0,0,1) corresponds to the $2^{nd}$ row/column … (1,1,1) corresponds to the $8^{th}$ row/column. Each number corresponds to the binary value of the tuple plus 1 (the right most number in each tuple is the least significant bit).

### 2.3.2 Controlled Not, Toffoli , Fredkin

Controlled not gates simply swap rows and columns of a density matrix. To determine which pairs of rows and columns to swap, we first find pairs of tuples which have control bits equal to 1, different values for target bits in the same location, and same values for all other bits. The pairs of numbers, corresponding to the binary values of the tuple pairs plus 1, are the pairs of rows and columns that need swapping.

For example, consider creating a controlled not gate on a system of 4 qubits with control qubit 1 and target qubit 2. Because the control bit already accounts for 1 of the tuple spaces, we will have $2^{(n-1)}$ tuples and thus $2^{(n-2)}$ tuple pairs. In each example below, since we have 4 qubit systems, we have four bits in each tuple. Also, tuples pairs are on top of each other.

1. The control bits are 1.  
   1 _ _ _, 1 _ _ _, 1 _ _ _, 1 _ _ _  
   1 _ _ _, 1 _ _ _, 1 _ _ _, 1 _ _ _
2. The target bits in tuple pairs have different values.  
   1 0 _ _, 1 0 _ _, 1 0 _ _, 1 0 _ _  
   1 1 _ _, 1 1 _ _, 1 1 _ _, 1 1 _ _
3. The remaining empty bit locations in the tuple pairs have the same values.  
   1 0 0 0, 1 0 0 1, 1 0 1 0, 1 0 1 1  
   1 1 0 0, 1 1 0 1, 1 1 1 0, 1 1 1 1
4. Find the numbers that correspond to each tuple.  
   1 0 0 0 = 9,   1 0 0 1 = 10,   1 0 1 0 = 11,   1 0 1 1 = 12  
   1 1 0 0 = 13,   1 1 0 1 = 14,   1 1 1 0 = 15,   1 1 1 1 = 16
5. The number pairs (9,13), (10, 14), (11, 15), and (12, 16) represent the rows and columns the controlled not gate algorithm swaps

Toffoli gates are similar to controlled not gates except that they have 2 control qubits. To determine which pairs of rows and columns to swap, we simply find pairs of tuples which have control bits equal to 1, different values for target bits in the same location, and same values for all other bits. The pair of numbers corresponding to the each tuple pair are the rows and columns that need swapping.

We will outline the steps for creating Toffoli gate on a system of 4 qubits with control qubits 2 and 3, and target qubit 4. Because the control bits already account for 2 of the tuple spaces, we will have $2^{(n-2)}$ tuples and thus $2^{(n-3)}$ tuple pairs.

1. The control bits are 1.   _ 1 1 _, _ 1 1 _
                             _ 1 1 _, _ 1 1 _
2. The target bits in the tuple pairs have different values.
                             _ 1 1 0, _ 1 1 0
                             _ 1 1 1, _ 1 1 1
3. The remaining empty bit locations in the tuple pairs have the same values.
                             0 1 1 0, 1 1 1 0
                             0 1 1 1, 1 1 1 1
4. Find the numbers that correspond to each tuple.
                             0 1 1 0 = 7, 1 1 1 0 = 15
                             0 1 1 1 = 8, 1 1 1 1 = 16
5. The number pairs (7,8) and (15, 16) represent the rows and columns the Toffoli gate swaps.

Fredkin gates are similar to controlled not gates except that they have 2 target qubits and an extra requirement. To determine which pairs of rows and columns to swap, we simply find pairs of tuples which have control bits equal to 1, different values for target bits in the same location, and same values for all other bits. In addition, we must also ensure that the target bits in the same tuple have different parity (that is, they cannot both be even or both be odd). The pair of numbers corresponding to the each tuple pair are the rows and columns that need swapping.

Now we will list the steps involved in creating a Fredkin gate on a system of 4 qubits with control qubit 3, and target qubits 1 and 4. Because the control bit accounts for 1 of the tuple spaces, and the extra requirement on the target bits accounts for another space, we will have $2^{(n-2)}$ tuples and thus $2^{(n-3)}$ tuple pairs.

1. The control qubits are 1.   _ _ 1 _, _ _ 1 _
                               _ _ 1 _, _ _ 1 _
2. The target bits in the same location in each tuple pair have different values but the target bits in the same tuple have different parity.
                               0 _ 1 1, 1 _ 1 0
                               1 _ 1 0, 0 _ 1 1
3. The remaining empty bit locations in the tuple pairs have the same values.
                               0 0 1 1, 1 1 1 0
                               1 0 1 0, 0 1 1 1

4.  Find the numbers that correspond to each tuple.
    $$0\ 0\ 1\ 1 = 4, \quad 1\ 1\ 1\ 0 = 15$$
    $$1\ 0\ 1\ 0 = 11, \quad 0\ 1\ 1\ 1 = 8$$
5.  The number pairs (4,11) and (8, 15) represent the rows and columns the Fredkin gate swaps.

### 2.3.3 Effectiveness

We compared the efficiency of the non-multiplying approach (where we simply swap rows and columns) versus the multiplying approach (where we create the gate matrices and multiplied them with the density matrix). We ran controlled not operations on `rand(2`$^n$`, 2`$^n$`)` matrices with target qubit 7 and control qubit 2. `Rand(x,y)` creates a x×y matrix filled with uniformly distributed random numbers on the interval (0,1).

Table 3: Comparison of non-multiplying approach versus multiplying approach

|  | 10 qubits |  | 12 qubits |  |
| --- | --- | --- | --- | --- |
| Number of gate operations | Time for non-multiplying (sec) | Time for multiplying (sec) | Time for non-multiplying (sec) | Time for multiplying (sec) |
| 1000 | 235 | 280 | 3020 | 5030 |

The non-multiplying approach shows a definite performance increase. However, the non-multiplying algorithms are very specific. They do each operation perfectly and thus cannot simulate operational errors. By building the matrix operations through tensoring, it is easier to add in small variances or other errors to achieve more realistic simulations.

### 3. Lognormal and Gaussian Distributed Operational Errors

We investigated the rate of degradation of different error distributions. In particular we looked at the differences between lognormal and Gaussian distributed operational errors on the Hadamard transform. A Hadamard operation is equivalent to a phase shift with $\phi = \pi$ followed by a rotation with $\theta = \pi/4$. We represent operational errors by adding small random numbers to the angles of the rotation and phase shift.

$$\begin{bmatrix} \cos(\theta) & -\sin(\theta) \\ \sin(\theta) & \cos(\theta) \end{bmatrix} \qquad \begin{bmatrix} 1 & 0 \\ 0 & e^{i\phi} \end{bmatrix}$$

Rotation(θ)                    Phase shift ( φ )

To set up our model, we used a standard Gaussian random number generator with $\mu = 0$ and $\sigma = (.1)^{.5}$ and a standard lognormal random number generator with $\mu = 0$ and $\sigma = .296$. These two parameters give the lognormal distribution a mean of 1.045 and a variance of .1. To center the lognormal distribution at 0, we simply subtracted 1.045 from each lognormally distributed random number.

Ideally, applying a Hadamard transform twice to the density matrix leaves the density matrix unchanged. We applied the Hadamard transform 40 times on a system of 3 qubits whose |000><000| term was initially 1 and recorded the |000><000| term after even numbered applications of the transform. We ran this setup 10,000 times using both Gaussian and lognormal random numbers centered at 0 with a variance of .1. After adding operational errors with a variance of .1, the |000><000| term showed an immediate, significant decrease.

Table 4: Hadamard Transform Degradation

|  | \|000> after 2 Hadamard transforms | \|000> after 4 Hadamard transforms | After 8 |
|---|---|---|---|
| Gaussian Variance = .1 | 0.5525 - 0.0000I | 0.3462 - 0.0000I | 0.2462 + 0.0000I |
| Lognormal Variance = .1 | 0.5761 - 0.0000I | 0.3706 - 0.0000i | 0.2689 - 0.0000I |

The results for the .1 variance case are statistically significant. We used a single sided, large sample test with $\alpha = .05$. The p-value for the hypothesis test that the Gaussian and lognormal values are equal is on the order of $10^{-8}$ and the power is approximately .9965. Gaussian distributed operational errors degrade qubits more quickly than lognormally distributed operational errors. Intuitively, one would expect this result since Gaussian distributions are two-tailed while lognormal distributions are only single-tailed.

### 4. 2-Qubit Multiplier Circuits

We have also examined the resistance of different 2-qubit multiplier circuits to operational errors. These circuits consist of Toffoli gates and controlled not gates and require 4 ancillary qubits. We wrote a program that exhaustively searches for circuits that implement desired function and give desired outputs. We found only 3 unique circuit layouts.

### 4.1 Circuit Schematics

X  Y

```
        *  W  Z
        •  •
     •  •
    ─────────
    3  2  1  0
```

The numbers 0, 1, 2, and 3 correspond to the ancillary qubits labeled 0, 1, 2 and 3. W, X, Y, and Z correspond to the input qubit registers W, X, Y, and Z. We can represent the input and output state by writing |X, Y, Z, W, 0, 1, 2, 3>. The schematics for the circuit layouts are shown below.

Figure 4: Solution 1

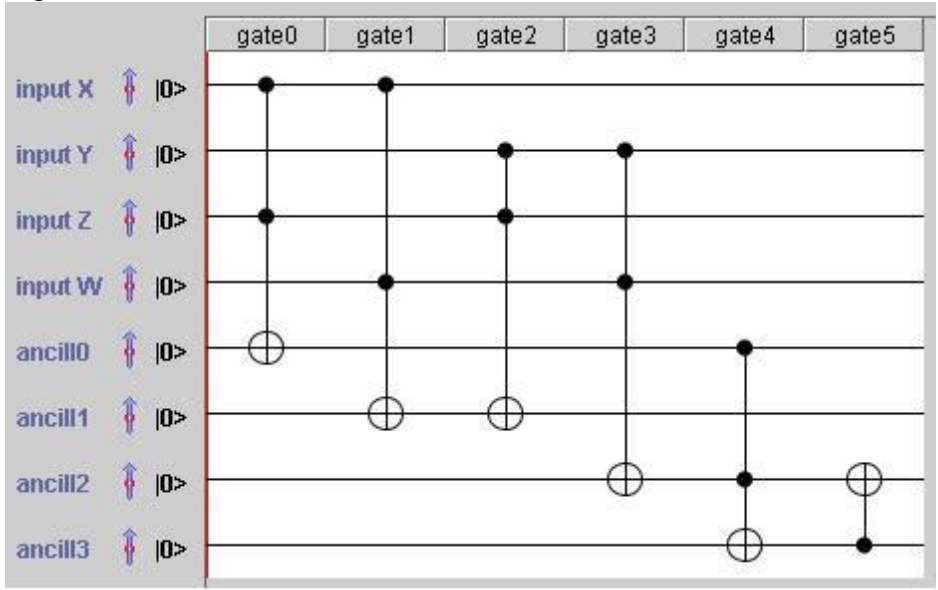

Figure 5: Solution 5

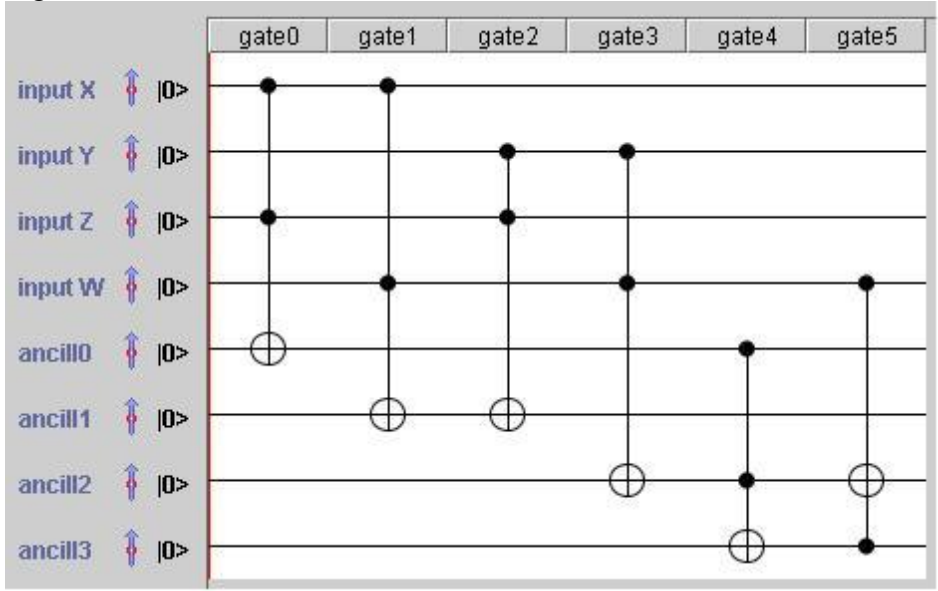

Figure 6: Solution 8

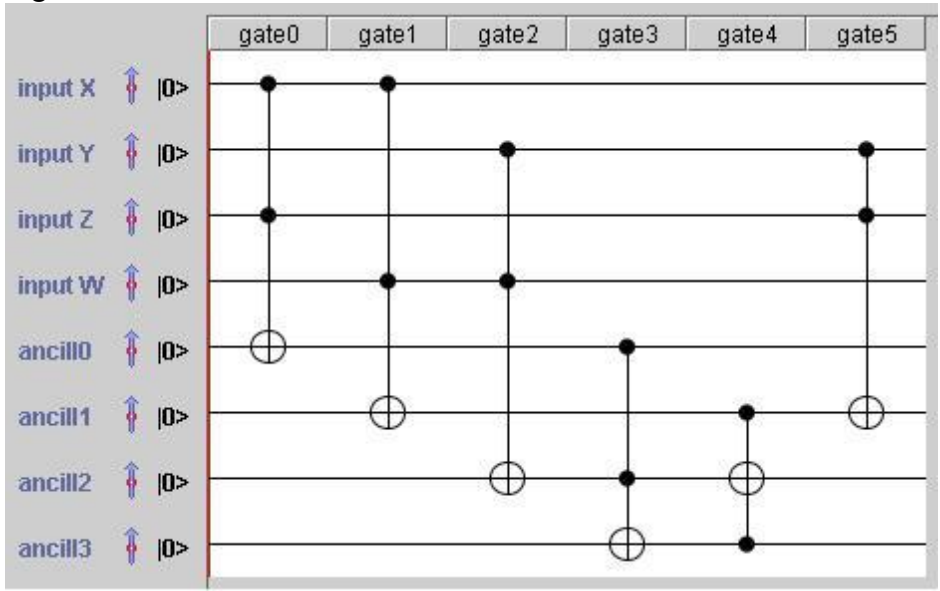

## 4.2 Experimental Setup

We tested the 2-qubit multiplier gates with the input X =1, Y=1, W= 0, Z =1, ancill0 = 0, ancill1 = 0, ancill2 = 0, and ancill3 = 0.  This input is |11100000> or a $2^8 \times 2^8$ density matrix with a 1 at (225, 225).   In an ideal operation, the output is X =1, Y=1, W= 0, Z =1, ancill1 = 1, ancill2 = 1, ancill2 = 0, and ancill3 = 0.   Written another way, |11101100>, or a $2^8 \times 2^8$ density matrix with a 1 at (237, 237).

Toffoli gates are basically controlled controlled not gates.  Not gates can be broken down into a phase shift with $\phi = \pi$ followed by a rotation with $\theta = \pi/2$.  We can then take those not gates and transform them into a controlled not gate or a controlled controlled not gate. To simulate operational errors, we add small Gaussian distributed random numbers to the angles θ and ϕ that create the not gate. We generate a new random number for each angle and each gate. Therefore the only operational errors are on the not gates (i.e. the target qubits), and we are not investigating correlated errors.

Table 5: 2-Bit Multiplier Circuitry Output

|            | Soln1     | Soln5    | Soln8    | Ideal |
|------------|-----------|----------|----------|-------|
| (225,225)  | 0.008078  | 0.00821  | 0.008198 | 0     |
| (229,225)  | -1E-04    | 2.93E-05 | -0.00012 | 0     |
| (233,225)  | 7.94E-05  | -9.8E-05 | 0.00027  | 0     |
| (237,225)  | 2.57E-05  | -0.00016 | 0.000145 | 0     |
| (225,229)  | -1E-04    | 2.93E-05 | -0.00012 | 0     |
| (229,229)  | 0.081716  | 0.08267  | 0.082597 | 0     |
| (233,229)  | 2.57E-05  | -0.00016 | 0.000145 | 0     |
| (237,229)  | -0.00041  | 0.000474 | 0.000766 | 0     |
| (225,233)  | 7.94E-05  | -9.8E-05 | 0.00027  | 0     |
| (229,233)  | 2.57E-05  | -0.00016 | 0.000145 | 0     |
| (233,233)  | 0.082003  | 0.082255 | 0.082163 | 0     |
| (237,233)  | 0.0001    | 0.000926 | 0.000101 | 0     |
| (225,237)  | 2.57E-05  | -0.00016 | 0.000145 | 0     |
| (229,237)  | -0.00041  | 0.000474 | 0.000766 | 0     |
| (233,237)  | 0.0001    | 0.000926 | 0.000101 | 0     |
| (237,237)  | 0.828203  | 0.826865 | 0.827042 | 1     |
|            |           |          |          |       |
| Trace distance | 0.171798 | 0.173141 | 0.172964 | 0 |
| Fidelity   | 0.910057  | 0.909321 | 0.909419 | 1     |

Note: All other locations in output matrices have a value of 0. The trace distance and fidelity were calculated with the average of 100,000 noisy circuit outputs, and the ideal output as inputs. These trace distances and fidelity values also show no statistically significant differences.

### 4.3 Circuit Analysis

To analyze differences between the 3 circuit layouts, we looked at 3 different values: the (237,237) term in the output, the trace distance between the noisy circuitry output and perfect circuitry output, and the fidelity between the noisy circuitry output and the perfect circuitry output. After running the experiment multiple times (collecting 80,000-100,000 samples for each of the 3 circuit layouts) and conducting one-way ANOVA tests on each mean, we found that the circuits showed no statistically significant differences. All the p-values, for the null hypothesis that the means are the same, were in the range of .27 to .46.

Despite the conclusion from the statistical analysis, the data from the multiple experiments seems to suggest that soln1 is slightly more robust than soln5 and soln8. Also, it seems that soln5 and soln8 show no significant differences between each other.

Because we only introduce errors on the target qubits, the input qubits always remain in perfect states. Also, there is no way for incorrect values to appear at ancill2 and ancill3 because all Toffoli gates and controlled not gates with target qubits ancill2 and ancill3 lack the required control qubits. Since errors are introduced only through the not gates on target qubits, ancill2 and ancill3 always remain 0. Thus, output values appear only at locations that correspond to |1110, ancill0, ancill1, 0, 0>. Possible state vectors corresponds to the locations (225, 225), (229, 229), (233, 233), and (237, 237) in the density matrix. Therefore, it is not a coincidence that these 4 locations have significantly large values in the output.

As one would expect, the (237, 237) term is largest despite the large variance of .1. A variance of .1 is unrealistic but it helps exaggerate the differences between the circuits. The (237, 237) term represents the probability of all the gates working correctly and outputting the correct answer, |11101100>, despite the operational errors.

The (225, 225) term represents the probability of none of the circuits working correctly. As a result, the output is the same as the input, |11100000>. The (229, 229) term corresponds to |11100100> and represents the probability of the circuit correctly operating on ancill1 but not changing the value in ancill0. Lastly, the (233, 233) term corresponds to |11101000> and represents the probability of the circuit correctly operating on ancill0 correctly but not changing the value in ancill1.

The (225, 225) term is smaller than the (229, 229) term and (233, 233) term because the (225, 225) term results from essentially 2 errors: not flipping the value in ancill0 and not flipping the value ancill1. The (229, 229) term and (233, 233) term result from the error of not changing one of the values. The likelihood of 2 errors is much less than the likelihood of just 1 error.

## 5. Conclusion

We developed the groundwork for a general quantum computer simulator in Matlab. We analyzed the efficiency of the basic matrix creation algorithms and gate algorithms. Also we showed that different error distributions have markedly different effects on a system. Lastly, we found that there is no statistical difference between the 3 different 2-qubit multiplier circuits in the presence of operational errors on the target qubits.

## 6. Acknowledgements

Thanks to Paul Black at NIST (National Institute of Standards and Technology) for his assistance and advice. Also I'd like to thank Alan Heckart for his statistical help and the quantum information / Bose-Einstein Condensation (QIBEC) group at NIST for their help and interesting ideas.

13Table 6: Times from Both Single Qubit Gate Algorithms

|   | Algorithm in Figure 2 (sec) | | | | Algorithm in Figure 1 (sec) | | |
|---|---|---|---|---|---|---|---|
|   | $1^{st}$ | $(n/2)^{th}$ | $n^{th}$ | | $1^{st}$ | $(n/2)^{th}$ | $n^{th}$ |
| 14 | 0.12891 | 0.9631 | 0.10688 | | 2.9 | 2.7 | 1.7701 |
| 15 | 2.8847 | 1.7089 | 2.1138 | | 6 | 5.7 | 3.931 |
| 16 | 4.7741 | 2.8701 | 3.7418 | | 12.3 | 11.7 | 7.694 |
| 17 | 9.3094 | 5.868 | 7.6063 | | 25.3 | 24.2 | 14.9465 |
| 18 | 17.5127 | 11.541 | 14.1138 | | 62.5 | 49.8 | 30.0846 |
| 19 | 35.799 | 22.816 | 29.1313 | | 138.1 | 106.5 | 75.6685 |
| 20 | 73.2862 | 44.7853 | 55.4327 | | 284.7 | 213.1 | 151.6205 |
| 21 | 147.3121 | 83.4493 | 105.3932 | | 505.2 | 464.3 | 309.8247 |
| 22 | 305.6424 | 170.1994 | 212.2553 | | 1109.3 | 1040.5 | 636.7049 |
| 23 | 619.6424 | 360.9448 | 421.2553 | | 2235.8 | 2097.7 | 1284.823 |

**References**

[1] Jumpei Niwa, Keiji Matsumoto, Hiroshi Imai. General-Purpose Parallel Simulator for Quantum Computing. [Quant-ph/0201042].